\documentclass[twocolumn,prl,amsmath,amssymb,amsfonts,superscriptaddress,floatfix,nopacs]{revtex4}
\usepackage{graphicx}
\usepackage{dcolumn}
\usepackage{bm}
\usepackage{rotating}
\usepackage{color}

\begin{document}
\newcommand{\red}{\color{red}}

\title{
The search for strong topological insulators.
}
\date{\today}
\author{M.~Klintenberg}
\thanks{Mattias.Klintenberg@fysik.uu.se}
\affiliation{
Department of Physics and Astronomy, Uppsala University,
Box 516, SE-75120, Uppsala, Sweden
 }

\maketitle
{\bf
Topological insulators 
\cite{Moore_2010,Hasan_2010,Kane_2008,Buttiker_2009,Moore_2009_a,Moore_2009_b} 
is a new quantum phase of matter with exotic properties such as dissipationless 
transport and protection against Anderson localization. 
\cite{Roushan_2009}
These new states of quantum matter could be one of the missing 
links for the realization of quantum computing \cite{Collins_2003,Bonderson_2010} 
and will probably result in new spintronic or magnetoelectric devices. 
Moreover, topological insulators will be a strong competitor with 
graphene 
in electronic application. Because of 
these potential application the topological insulator research has 
literally exploded during the last year. Motivated by the fact that 
up-to-date only few 3D systems are identified to belong to this new 
quantum phase 
\cite{Lin_2010,Fu_2007,Moore_2007,Roy_2009,Hsieh_2008,Hsieh_2009,Xia_2009,Zhang_2009_b,Chen_2009} 
we have used massive computing 
in combination with data-mining to search for new strong topological 
insulators. In this letter we present a number of non-layered 
compounds that show band inversion at the $\Gamma$-point, a clear 
signal of a strong topological insulator.  
}

The prediction and realization of the 3D topological insulators 
\cite{Lin_2010,Fu_2007,Moore_2007,Roy_2009,Hsieh_2008,Hsieh_2009,Xia_2009,Zhang_2009_b,Chen_2009} 
grew out of the work on the quantum Hall effect and quantum spin Hall effect in 2D 
\cite{Murakami_2004,Kane_2005_a,Kane_2005_b} and 
the essence of these central articles 
\cite{Fu_2007,Moore_2007,Roy_2009,Hsieh_2008,Hsieh_2009,Xia_2009,Zhang_2009_b,Chen_2009,Murakami_2004,Kane_2005_a,Kane_2005_b} 
is captured in a number letters describing topological insulators 
\cite{Kane_2008,Buttiker_2009,Moore_2009_a,Moore_2009_b,Moore_2010}.

In school we have been taught that materials can be divided into 
insulators and conductors. Lately a new state of matter has emerged, 
the so-called strong topological insulators in 3D. These materials 
can be viewed as insulators in the bulk but with metallic surface 
states, resembling the Dirac cones in graphene. These metallic 
surface states exhibit remarkable properties such as dissipation 
less transport and topological protection against perturbations and 
impurity scattering. Not surprisingly these new materials have 
spawned great attention and research in the science community 
because of the potential of interesting application. 

An insulator is characterized by its non-ability to conduct, 
{\it id est} an external electric field does not cause current 
flow. Alternatively we think of an insulator as a system 
that lacks low lying excited states and it takes a finite amount of 
energy for an electronic excitation, {\it i.e.} we have an electronic band 
gap. It appears that the electronic band gap is the one characteristic 
needed to characterize an insulator and this was true until the quantum 
Hall effect (QHE) came around. The quantum Hall state (QHS) 
shows up when, for example, a two-dimensional electron gas is subject 
to a magnetic field. The interior of the QHS has a band gap but 
contrary to a ”normal” insulator the QHS has edge states 
that do not show a band gap, {\it i.e.} there is a current on the boundary 
of the sample. This very special charge flow is one directional thus 
making it protected against scattering \cite{Hasan_2010}. Both the insulating- 
and QHS systems show bulk band gaps but the latter also 
have the surface states that are insensitive to scattering. So what 
is the difference between the two systems? It turns out that it is the 
topology of the occupied bands that does it, or more specifically, the 
topological class of the bulk band structure. 

Next came the quantum spin Hall effect (QSHE) \cite{Kane_2005_a,Kane_2005_b}. 
The quantum spin Hall state (QSHS) is characterized by yet another 
topological invariant (knot) and can exist without a magnetic field 
(a QHS required a magnetic field) thanks to the spin-orbit interaction. 
The bulk has a band gap where as the edge states do not. In graphene these edge 
states are spin-filtered thus making the different spins travel in 
opposite directions. The QSHE can give us edge states without a magnetic field in 2D thanks 
to the spin-orbit interaction. But what happens when we go 3D? In refs. 
\cite{Fu_2007,Moore_2007,Roy_2009,Hsieh_2008,Hsieh_2009,Xia_2009,Zhang_2009_b,Chen_2009} 
it was both predicted and experimentally verified that 
topological insulators do exist in 3D and it is the spin-orbit 
interaction that helps out. We can think of these edge states in 3D 
topological insulators as a higher dimensional version of the 1D edge 
states in the QSHS. In 3D the simplest form of edge state can be viewed 
as a Dirac fermion metal with a linear excitation energy, {\it c.f.} the 
energy spectrum of mass less Dirac fermions (Dirac cone). Moreover, 
because these surface states are topologically protected by 
time-reversal symmetry, Anderson localization does not occur even under 
strong disorder. A topological insulator is simply not allowed to lose 
the metallic surface state, {\it id est} become gaped or localize. This 
was recently experimentally verified for Bi$_{x}$Sb$_{1-x}$ \cite{Roushan_2009}.

\begin{figure}[ht]
 \begin{center}
\includegraphics[bb=120 340 600 580, angle=-90,width=30mm]{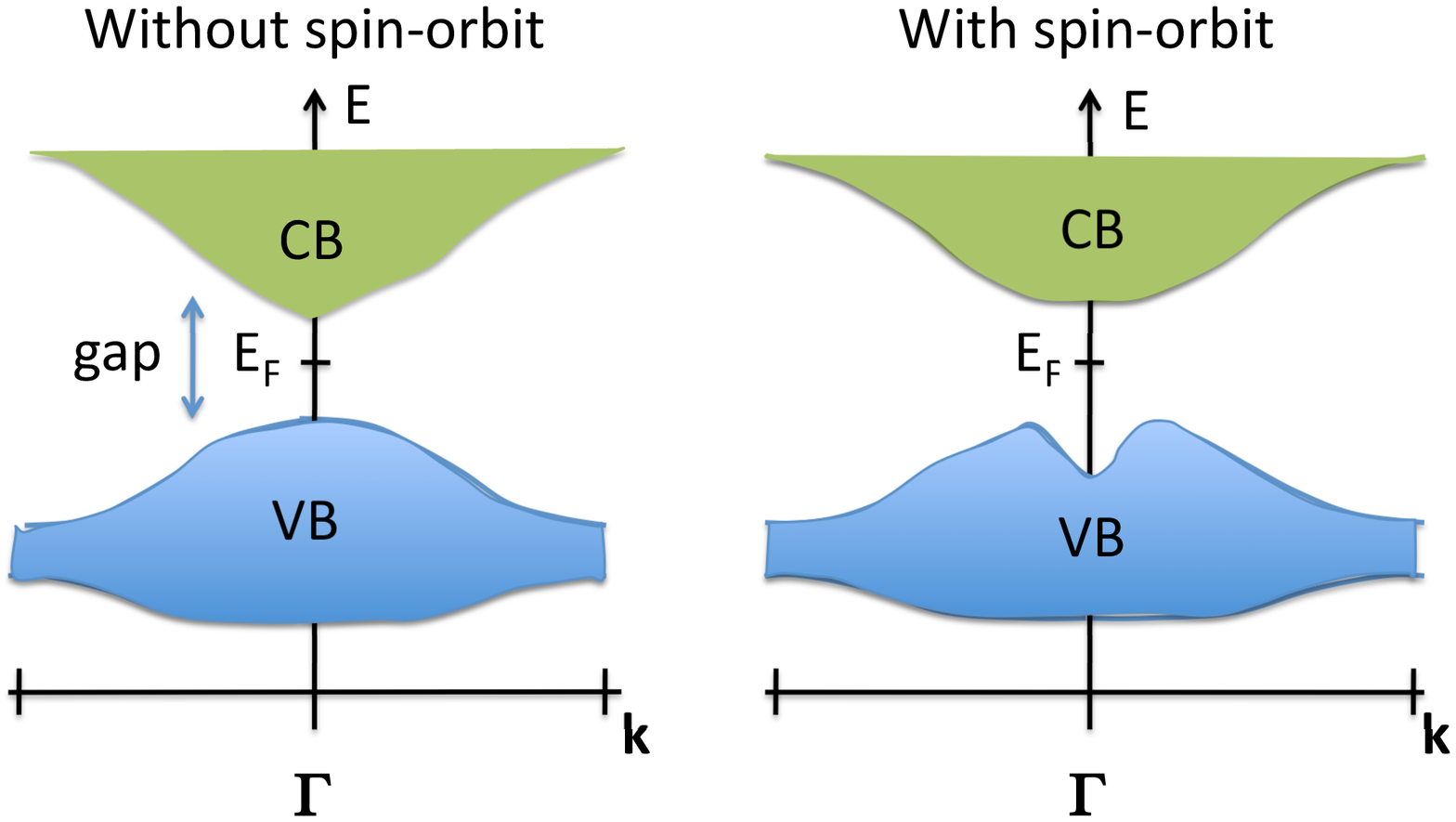}
\caption{\label{band_inv} Band inversion because of strong spin-orbit coupling 
in small electronic band gap materials can be identified by searching for the 
anti-crossing feature around the $\Gamma$-point as the spin-orbit coupling 
is included in the calculation (right panel), {\it c.f.} without spin-orbit 
interaction (left panel).\\
}
 \end{center}
\end{figure}

When data-mining electronic structures in the search for new strong 
topological insulators the most straight forward method is to 
identify materials with a negative band gap \cite{Zhang_2009_b}. To put it differently 
the electronic structure should develop an anti-crossing feature 
at the $\Gamma$-point when the spin-orbit interaction is turned 
on as compared to no spin-orbit coupling. A typical 
band-inversion is shown in Figure \ref{band_inv}. Thus the data-mining 
algorithm is instructed to search for small gaped materials with 
an anti-crossing feature at the $\Gamma$-point. Note that the data-mine 
of electronic structures have been calculated with spin-orbit 
coupling.

To show proof-of-concept the first three known second generation 
topological insulators Bi$_2$Te$_3$, Bi$_2$Se$_3$ and Sb$_2$Te$_3$ 
\cite{Xia_2009,Zhang_2009_b,Chen_2009} were 
investigated for band-inversion when spin-orbit interaction was 
turned on/off. Indeed all three materials develop the characteristic 
anti-crossing feature as spin-orbit coupling is included in the 
electronic structure calculation.

\begin{figure}[ht]
 \begin{center}
\includegraphics[bb=40 340 600 580, angle=-90,width=30mm]{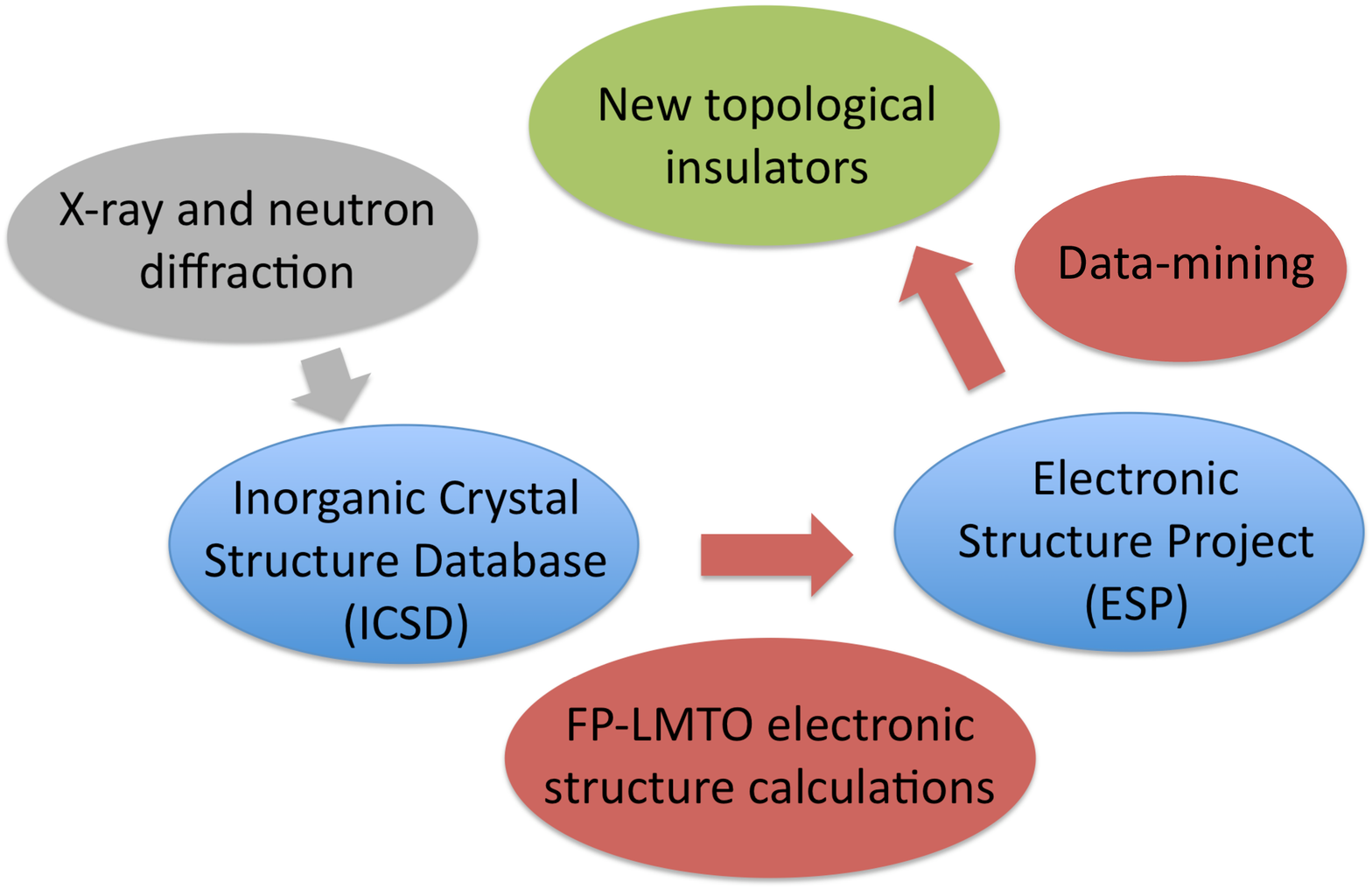}
\caption{\label{data_mine} A schematic description of the search for new 
topological insulators. The key ingredient is the electronic structure project 
(ESP) that contain the electronic structures of tens of thousands of inorganic 
compounds. Data mining algorithms on the ESP has the potential to identify a 
variety of new functional materials.\\
}
 \end{center}
\end{figure}

Before presenting results, the data-mining of electronic structures 
is briefly explained. The complete process is shown in Figure 
\ref{data_mine}. The starting point is the inorganic crystal 
structure database (ICSD) 
which is a collection of some 
130,000 experimental crystal structures obtained by X-ray and neutron 
diffraction. Using only the crystal structure as input we 
have calculated the electronic structure for about 60,000 entries 
in the ICSD using a full-potential linear muffin-tin orbital implementation 
(FP-LMTO) \cite{Skriver_1984} of density functional theory (DFT) within 
the local density approximation (LDA). The obtained library is 
maintained within the electronic structure project (ESP) 
and has been made available on the web \cite{espweb}. The 
data-mining process in the search for new strong topological 
insulators is rather straight forward. Non-layered small gap 
materials ($\le$0.5eV) with an anti-crossing feature at the 
$\Gamma$-point are identified. Furthermore it is verified that the 
anti-crossing feature disappears when the spin-orbit interaction 
is switched off. The data-mining identifies 17 compounds as potential 
strong topological insulators and these are presented in Table \ref{tab1}.
Note that because the ESP take crystal structures from the ICSD all 
compounds identified in this study exist and have been structurally 
determined.

\begin{table}
\title{Supplementary information 2}\maketitle
  \begin{tabular}{llll}
  \hline
  Material & Spgrp & Struct. & LDA      \\
           &       & type    & band gap [eV]\\
  \hline
  Ca$_3$PbO & P m -3 m  & CaTiO$_3$ & 0.2\\
  Sr$_3$PbO & P m -3 m  & CaTiO$_3$ & 0.1\\
  Ba$_3$PbO & P m -3 m  & CaTiO$_3$ & 0.1\\
  Yb$_3$PbO & P m -3 m  & CaTiO$_3$ & 0.2\\
  Ca$_3$SnO & P m -3 m  & CaTiO$_3$ & 0.2\\
  Sr$_3$SnO & P m -3 m  & CaTiO$_3$ & 0.1\\
  Yb$_3$SnO & P m -3 m  & CaTiO$_3$ & 0.1\\
  GdPtSb & F -4 3 m & AlLiSi & 0.2\\
  Bi$_2$SeTe$_2$ & R -3 m H & Bi$_2$Te$_3$ & 0.3\\
  Bi$_2$STe$_2$ & R -3 m H & Bi$_2$Te$_3$ & 0.3\\
  PbTl$_4$Te$_3$ & I 4/m c m & In$_5$Bi$_3$ & 0.1\\
  BiTl$_9$Te$_6$ & I 4/m c m & In$_5$Bi$_3$ & 0.1\\
  BiTlTe$_2$ & R -3 m H & NaCrS$_2$ & 0.0\footnote{The material has small hole pockets.}\\
  SbTlTe$_2$ & R -3 m H & NaCrS$_2$ & 0.2\\

  Bi$_2$TeI & C 1 2/m 1 & Bi$_2$TeI & 0.1\\
  GeSb$_4$Te$_7$ & P -3 m 1 & AgBiSe$_2$& 0.2\\
  HgKSb & P 63/m m c & KZnAs & 0.2\\
  \hline
\end{tabular}
\caption{Results of the mining algorithm for second generation non-trivial topological insulators. 
17 compounds are identified as new potential topological insulators. ARPES measurements should confirm our 
findings. The electronic band structures, partial and total density of states as well as references to 
to the crystal structure determinations are given for all materials in the supplementary information.}
\label{tab1}
\end{table}

\begin{figure}[ht]
 \begin{center}
\includegraphics[angle=-90,width=80mm]{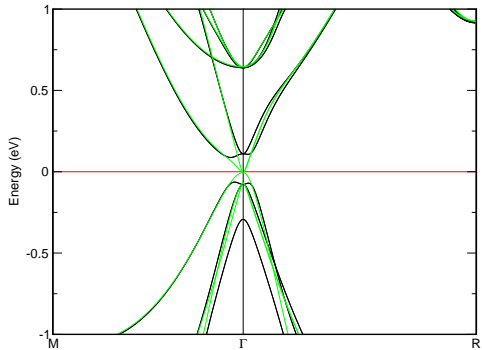}
\caption{\label{fig3} Ca$_3$OSn is an example of inverse perovskite that show band inversion. The 
material transforms from direct gap (no spin-orbit interaction included, green line) at the $\Gamma$-point to 
in-direct gap (around the same point in inverse space) when the spin-orbit is included (black line) in the 
electronic structure calculation, {\it i.e.} the upper most valence band is pushed way and the 
anti-crossing feature appears. The LDA band gap is 0.2 eV which qualifies the compound for 
operation under ambient conditions.\\
}
 \end{center}
\end{figure}

The first group of materials in Table \ref{tab1} belong to the CaTiO$_3$ (perovskite) 
structure type but are inverse-perovskites. Perovskites are interesting materials because 
these host a variety of materials properties such as superconductivity, spin dependent 
transport, ferroelectricity and colossal magnetoresistance. Present calculations indicate 
that the strong topological insulator phase now can be added to this list. Through out the 
Brillouin zone these materials have an in-direct LDA band gap ranging from 0.1 to 0.2 eV. The 
characteristic anti-crossing feature is present at the $\Gamma$-point and if the spin orbit 
coupling constant is set to zero the anti-crossing feature disappears. From the partial 
density of sates we conclude that these bands have mainly $p$-character. The electronic 
band structure for Ca$_3$OSn is shown in Figure \ref{fig3}.

Next in Table \ref{tab1} is the half-Heusler compound GdPtSb with a  AlLiSi structure type. 
The half Heusler family has been discussed in \cite{Lin_2010} but there with a distorted structure. 
Here it is explicitly demonstrated for GdPtSb that with the experimental structure the 
material develops an anti-crossing feature when the spin-orbit term is included in the 
calculations and we speculate that this is also true for the other materials in \cite{Lin_2010}. 
GdPtSb has an indirect band gap of 0.2 eV. 

\begin{figure}[ht]
 \begin{center}
\includegraphics[angle=-90,width=80mm]{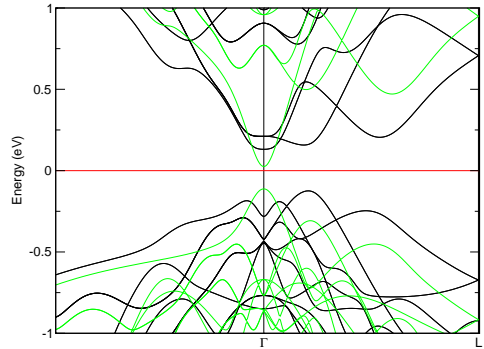}
\caption{\label{fig4} Bi$_2$SeTe$_2$ has the same structure type as the celebrated 
topological insulator Bi$_2$Te$_3$. We see the same effect of the spin-orbit interaction 
as in the Ca$_3$OSn case. The valence band at the $\Gamma$-point is pushed away and the anti-crossing 
feature show up when the spin-orbit interaction is included (black line, green line for no spin-orbit interaction). 
This compound could work under ambient 
conditions because of the 0.3 eV band gap.\\
}
 \end{center}
\end{figure}

Bi$_2$SeTe$_2$ and Bi$_2$STe$_2$ both have the Bi$_2$Te$_3$ structure type. In fact these 
materials can be viewed as Bi$_2$Te$_3$ but with every third Te replaced with Se and S, 
respectively. Bi$_2$Te$_3$ is a celebrated second generation topological insulator \cite{Zhang_2009_b} 
and both the Se and S variations have similar electronic structure to Bi$_2$Te$_3$ and 
show the anti-crossing feature at the $\Gamma$-point when spin-orbit interaction is removed. 
The indirect band gaps are 0.3 eV, respectively. The electronic band structure 
for Bi$_2$STe$_2$ is shown in Figure \ref{fig4}. 
For Bi$_2$SeTe$_2$ the electronic band gap closes without the spin-orbit term and the 
material appear to be a topological metal within the local density approximation (LDA). 
However, LDA is known to underestimate the electronic band gap and using higher order 
corrections such as the GWA \cite{Kotani} could very well open up a band gap thus making 
it a topological insulator. We speculate that there are many not yet discovered topological 
metals, {\it i.e.} a bulk metal but with topologically protected surface states, and a 
data-mining search for these materials will be the subject for our next study.

PbTl$_4$Te$_3$ and BiTl$_9$Te$_6$ have a body-centered tetragonal Bravais lattice and 
belong to the In$_5$Bi$_3$ structure type family. In$_5$Bi$_3$ exhibit interesting 
electronic properties such as superconductivity. The two compounds both show the anti-crossing 
feature as spin-orbit is included and we note the curvature of these bands is smaller 
compared to, for example, the Bi$_2$Te$_3$ family. The indirect band gap is 0.1 eV, 
respectively.

SbTlTe$_2$ and BiTlTe$_2$ have the NaCrS$_2$ structure type. The former material has an indirect 
band gap of 0.2 eV. The latter electronic structure is more complex. Using LDA the material 
show small electron and hole pocket but could very well turn out to have a small band gap 
when investigated using higher order theory such as GWA. 

The last three materials in Table \ref{tab1} belong to different structure types and all have 
indirect band gaps but with more exotic electronic band structures. Bi$_2$TeI develop the 
anti-crossing feature at two high symmetry points ($\Gamma$ and V) in the Brillouin zone. 
GeSb$_4$Te$_7$ and HgKSb show the band inversion at the A-point. The latter also display a 
small hole pocket away from the A-point. 

To conclude we have used massive computing and data mining to search for and identify new potential 
topological insulators. In Table \ref{tab1} 17 new materials have been identified that all 
show an anti-crossing feature at the $\Gamma$-point (V, A for the last three) as the 
spin-orbit coupling is included. This feature is a clear signal for a topological insulator 
\cite{Zhang_2009_b}. As pointed out in reference \cite{Lin_2010} it is crucial to identify 
new groups of materials (electronic band structures) that are topologically non-trivial 
to maximize the likelihood for new physics as well as new devices 
\cite{Fu_2009,Qi_2009,Essin_2009,Dzero_2010,Linder_2010,Ran_2010,Bonderson_2010}.

\section{Methods}
The more then 60,000 electronic structures used in the present study were calculated 
using a highly accurate full-potential linear muffin-tin orbital (FP-LMTO) \cite{Skriver_1984} 
implementation of density functional theory (DFT) 
within the local density approximation (LDA) 
In essence DFT reduces the many-body problem of solving the electronic structure of some $10^{23}$ 
interacting electrons into to a one-body problem, thus making the calculation possible. 
A more detailed description can be found in \cite{espweb,Skriver_1984}.

\section{Acknowledgments}
The Swedish Research Council (VR) and the G\"oran Gustafsson Stiftelse are acknowledged for financial support.

\section{Additional information}
The authors declare no competing financial interests.

\end{document}